\documentstyle[prl,aps,floats,epsf]{revtex}

\begin{document}
\draft

\twocolumn[\hsize\textwidth\columnwidth\hsize\csname
@twocolumnfalse\endcsname

\title{
Phase transitions, stability, and  dielectric response of the 
domain structure in ferroelectric-ferroelastic thin films
}
\author{A.M. Bratkovsky$^{1}$ and A.P. Levanyuk$^{1,2}$}

\address{$^{1}$Hewlett-Packard Laboratories, 1501 Page Mill Road, Palo
Alto, California 94304\\
$^{2}$Departamento de F\'{i}sica de la Materia Condensada, C-III,
Universidad Aut\'{o}noma de Madrid, 28049 Madrid, Spain
}
\date{October 3, 2000}
\maketitle

\begin{abstract}

We present the first analytical study of phase transitions in
ferroelastic-ferroelectric epitaxial thin films on
exactly solvable model. 
The emerging domain structure with domains of equal width
(which may be  exponentially large on a ``soft" substrate)
always remains stable irrespective of the film thickness. 
The dielectric response  of an epitaxial film is {\em smaller} 
than that of a free film, in
striking contrast with assertions in the literature.


\end{abstract}
\vskip 2pc ]

Formation of ferroelastic domains in thin films due to elastic misfit
between the film and the substrate was predicted by Roytburd \cite
{Roytburd76}. The domain patterns are the focus of extensive studies,
especially in epitaxial films of perovskite ferroelectrics that are improper
ferroelastics\cite{see ref}. However, the main results in this field are
obtained by numerical methods or by making seemingly reasonable yet
unjustifiable approximations. Additionally, all previous results refer to
the case of equal elastic moduli of the film and the substrate. As a result,
the physics of the epitaxial ferroelectric-ferroelastic films remains
obscure, and this has led to generally accepted yet erroneous statements
in the literature.

In this paper we consider an exactly solvable case of a ferroelectric-proper
ferroelastic thin film described by a one-component order parameter, which
is either a strain tensor component or a polarization component \cite
{Strukov} exhibiting a second order phase transition. The model allows an
analytical treatment at all temperatures, including that in the vicinity of
the transition.

We shall begin with the case of the film in the absence of an external
electric field. The film is assumed to be perpendicular to the $z$-axis and
the order parameter the $u_{xy}$ component of the strain tensor. It is
assumed to be attached to an elastically isotropic substrate with the shear
modulus $\mu $, Fig.~1. The elastic moduli of the ferroelastic are supposed
to be the same as in the substrate with the exception of the ``soft''
modulus corresponding to the $u_{xy}$ component of strain. Thus, the Landau
thermodynamic potential has the form 
\begin{eqnarray}
F &=&\int dV[2Au_{xy}^{2}+2D\left( \nabla u_{xy}\right) ^{2}+Bu_{xy}^{4} 
\nonumber \\
&&+\mu \left( u_{ik}^{2}-2u_{xy}^{2}\right) ]  \label{eq1}
\end{eqnarray}
where $A=\alpha \left( T-T_{c}\right) $, with $\alpha ,D,\mu $ positive
constants, and we have only kept the most relevant terms. The equation of
state is 
\begin{equation}
\sigma _{xy}=2(A-D\nabla ^{2})u_{xy}+2Bu_{xy}^{3}.  \label{eq2}
\end{equation}

We first consider the loss of stability of the symmetric (paraelectric,
paraelastic) phase at the phase transition. According to standard procedure,
it corresponds to the first appearance of a non-trivial solution to the {\em %
linearized} equations of state. We shall look for the non-trivial solution
for the $x$-component of the displacement vector $u_{x}\equiv u(y,z).$ One
can swap the $x$ and $y$ axes to consider the $u_{y}$ strains instead. This
will only be the consideration of the inhomogeneous part of the strain. The 
{\em homogeneous} strain of the whole sample defines the change of its
volume and shape and is described by six independent components of the
(homogeneous)\ strain tensor. In our case, there is no loss of stability
with respect to homogeneous deformation since it would cost an infinite
elastic energy to produce such a strain with an infinitely thick substrate.

To find the inhomogeneous part of the strain $u_{x}$ at the phase transition
one should satisfy the equations of local equilibrium, $\partial \sigma
_{ik}/\partial x_{k}=0,$ which in the present case read 
\begin{equation}
\frac{\partial \sigma _{xy}}{\partial y}+\frac{\partial \sigma _{xz}}{%
\partial z}=0.  \label{eq3}
\end{equation}
We shall use the Fourier representation 
\begin{equation}
u(y,z)=\int u_{k}\left( z\right) \exp \left( iky\right) dk  \label{eq4}
\end{equation}
and find the first appearance of the non-trivial solution for $u$ for a
given  wave vector $k$. We then determine the $k$ where the
instability sets 
in first, and this will be the point of the stability loss of the symmetric
phase.

We obtain the following equations for the strain with the use of Eqs.(\ref
{eq3}),(\ref{eq2}) 
\begin{eqnarray}
\frac{d^{2}u_{k}}{dz^{2}}-\frac{A_k }{\mu }k^{2}u_{k} &=&0,\quad 0<z<l;
\label{eq:film} \\
\frac{d^{2}u_{k}}{dz^{2}}-k^{2}u_{k} &=&0,\quad -\infty <z<0,  \label{eq:sub}
\end{eqnarray}
where $A_k =A+Dk^{2}.$ At the free surface $\left( z=l\right) $ the boundary
condition reads $\sigma _{xz}\left( l\right) =0,$ which is equivalent to $%
du_{k}(z)/dz=0$. In addition, the displacement $u_{k}\left( z\right) $ and
the stress $\sigma _{xz}\left( z\right) $ should be continuous at the
interface $z=0,$ and the stress should vanish at $z\rightarrow -\infty .$
\begin{figure}[t]
\epsfxsize=3.2in \epsffile{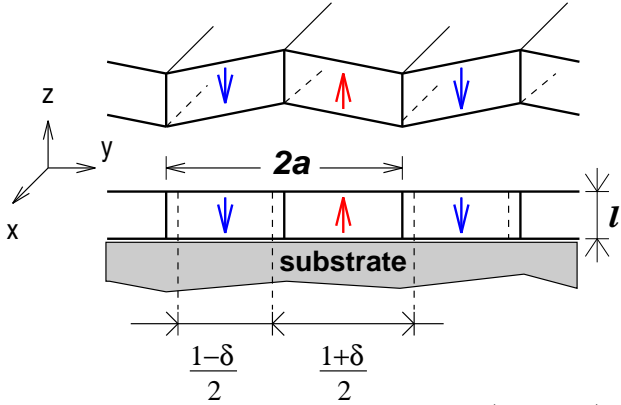}
\caption{Schematic of the domain structure (top panel) with a period $2a$
in a thin ferroelastic film of thickness $l$ attached to a substrate
(bottom panel). The relative displacement of the domain walls by $\protect%
\delta $ would change the homogeneous strain in the bulk of the film. Since
the appearance of the homogeneous deformation in the film is very costly,
the change in the domain width $\protect\delta $ in the field will be
suppressed, and the dielectric response would {\em not} be enhanced in
comparison with a monodomain free-standing film. }
\label{fig:fig1}
\end{figure}

Let us first consider the case of $A_k <0$, which would lead to a loss of
stability of the paraphase. The solution of Eqs. (\ref{eq:film}), (\ref
{eq:sub}) is

\begin{eqnarray}
u_{k}(z) &=&F\cos \eta k(z-l),\quad 0<z<l;  \label{eq8} \\
u_{k}(z) &=&G\exp |k|z,\quad -\infty <z<0,
\end{eqnarray}
where $\eta^{2}=-A_k/\mu .$ The boundary conditions give us the condition of
the existence of the non-trivial solutions 
\begin{equation}
\cot \eta kl=\eta.  \label{eq10}
\end{equation}
This equation does have a solution for $A_k<0,$ while there is no solution
at $A_k >0$, hence the loss of stability takes place for $A_k  <0.$ For the
region of interest $\left( \eta\ll 1\right) $ the approximate solution is $%
\eta \simeq \pi /2kl$ or

\begin{equation}
\mid A\mid -Dk^{2}=\frac{\pi ^{2}\mu }{4k^{2}l^{2}}  \label{eq11}
\end{equation}

The minimum value $\mid A\mid _{c}$ corresponds to 
\begin{equation}
k_{m}=(\frac{\pi }{2})^{1/2}\frac{\mu ^{1/4}}{D^{1/4}l^{1/2}}\sim \frac{1}{%
d_{at}^{1/2}l^{1/2}}  \label{eq12}
\end{equation}
where usually (not on very ``soft'' substrates) $\left( D/\mu \right)
^{1/2}\sim d_{at},$ with $d_{at}$ the interatomic distance (see, e.g. \cite
{Strukov}). This means that the loss of stability of the symmetric phase
takes place at 
\begin{equation}
\mid A\mid _{c}=\frac{\pi D^{1/2}\mu ^{1/2}}{l}\sim \mu \frac{d_{at}}{l},
\label{eq13}
\end{equation}
or at 
\begin{equation}
T=T_{c1}=T_{c} - \frac{\pi D^{1/2}\mu ^{1/2}}{\alpha l}.  \label{eq14}
\end{equation}
The temperature of the phase transition is lowered by about $T_{at}\frac{d}{l%
},$ where $T_{at}$ is the characteristic ``atomic'' temperature. Since
normally $T_{at}\sim \left( 10^{2}-10^{3}\right) T_{c},$ one may expect a
complete suppression of the second order transition for films with
thicknesses of hundreds of atomic layers depending on the particular
materials parameters. The inhomogeneous structure that forms when the system
looses stability is a domain structure close to the phase transition, where
the widths of the domain walls and the domains themselves are comparable
(see, e.g. \cite{Chensky}). One can readily see that this is valid in the
present case since the domain wall width $W=\left( D/2\mid A\mid \right)
^{1/2}$.

We consider next the domain structure not very close to the phase transition
in a state with the spontaneous strain $u_{xy}^{0}$. There the domain wall
width is much smaller than the width of the domains and one can use the
linearized equation of state, obtained by expanding the free energy (\ref
{eq1}) about the spontaneous deformation, 
\begin{eqnarray}
\sigma _{xy} &=&2M(u_{xy}-u_{xy}^{0}),  \label{eq15} \\
u_{xy}^{0} &\equiv &u_{0}=\pm \left( -A/B\right) ^{1/2},  \label{eq:uspont}
\end{eqnarray}
where $M\equiv -2A$ is $\ll \mu $ when the system is close to the transition
(``soft'' modulus), with the gradient term being the origin of the domain
wall energy.

Initially we shall assume that all the domains have the same width, which we
will find by minimizing the sum of the elastic energy and the (surface)\
energy of the domain walls. We shall follow the standard procedure
(see e.g. \cite{Khachaturyan}). Firstly, we consider the film without
a contact to the 
substrate where we create a distribution of the spontaneous strain (a domain
structure) in such a way that there will be no stresses in the film. The
total length of the contact area of the film remains {\em exactly} the same
as if it were in the symmetric phase, Fig.~1. Then we join the film and the
substrate, maintaining the continuity of the displacements at the interface.
After the contact is made, the {\em inhomogeneous} stresses appear both in
the film and in the substrate, but no uniform stresses.

We consider a stripe-like domain structure with the spontaneous strain 
\begin{eqnarray}
u_{xy}^{0}(y,z) &=&u_{0},\hspace{.1in}0<z<l,\hspace{.1in}\left(
2n-1\right) a<y<2na;  \label{eq16} \\
u_{xy}^{0}(y,z) &=&-u_{0},\hspace{.1in}0<z<l,\hspace{.1in}2na<y<\left(
2n+1\right) a,  \nonumber
\end{eqnarray}
with the period $2a.$ There would be no stresses in the structure if $%
u_{0}^{2}=-A/B.$ We have to find the displacements appearing after the film
is attached to the substrate, $u_{x}(y,z)\equiv u\left( y,z\right) $. For
the film $\left( 0<z<l\right) $ Eq.(\ref{eq3}) takes the form 
\begin{equation}
M\frac{\partial ^{2}u}{\partial y^{2}}+\mu \frac{\partial ^{2}u}{\partial
z^{2}}=2M\frac{\partial ^{2}u_{xy}^{0}}{\partial y^{2}}.  \label{eq17}
\end{equation}
Since the domain pattern is periodic, the elastic displacements may be
represented as a Fourier series 
\begin{equation}
u\left( y,z\right) =\sum_{k}u_{k}\left( z\right) \exp \left( iky\right) ,%
\text{\qquad }k=\frac{\pi n}{a},  \label{eq18}
\end{equation}
where $n=\pm 1,\pm 2,\ldots $ After solving the resulting ordinary
differential equations with the above conditions one finds the elastic
energy by e.g. using the formula \cite{Mura} 
\begin{equation}
F_{el}=-\frac{1}{2}\int \sigma _{ij}u_{ij}^{0}dV  \label{eq19}
\end{equation}
with the result for the elastic (stray) energy per unit area of the film$:$%
\begin{equation}
\frac{F_{stray}}{{\cal A}}=\frac{16\mu u_{0}^{2}a}{\pi ^{3}}\eta
\sum_{j=0}^{\infty }\frac{1}{\left( 2j+1\right) ^{3}}\frac{1}{\coth \eta
kl+\eta },  \label{eq20}
\end{equation}
where $\eta =\sqrt{M/\mu }$. We shall first consider temperatures not very
far from the transition, where $\eta \ll 1.$ Since $\coth x>1$ for any
argument, one can then omit the second term in the dominator. We shall
suppose that the equilibrium period of the domain structure satisfies the
condition $\eta kl\gg 1$ and check later that this condition is fulfilled.
We can then put $\coth \eta kl\simeq 1$ for all terms in (\ref{eq20}) and
find the energy of the domain structure 
\begin{equation}
\frac{F}{{\cal A}}=\frac{\gamma l}{a}+\frac{14\zeta \left( 3\right) }{\pi
^{3}}\mu u_{0}^{2}\eta a,  \label{eq22}
\end{equation}
where the first term is the energy of the domain walls, with $\gamma $ the
domain wall surface energy. Therefore, the equilibrium period of the domain
structure is 
\begin{equation}
a=\left( \frac{\pi ^{3}}{14\zeta \left( 3\right) }\frac{\gamma l}{\mu
u_{0}^{2}\eta }\right) ^{1/2}=\left( \frac{4\pi ^{3}}{21\zeta \left(
3\right) }\frac{D^{1/2}l}{\mu ^{1/2}}\right) ^{1/2},  \label{eq23}
\end{equation}
since $\gamma =8\sqrt{2}D^{1/2}\mid A\mid ^{3/2}/3B$\cite{Strukov}. We must
check now if the above assumption $\pi \eta l/a\gg 1$ is fulfilled together
with $\eta \ll 1$. Together those constraints read 
\begin{equation}
\frac{\left( D/\mu \right) ^{1/2}}{l}\ll \frac{\mid A\mid }{\mu }\ll 1
\label{eq26}
\end{equation}
Since usually $\left( D/\mu \right) ^{1/2}\sim d_{at}\ll l,$ the condition
is satisfied not very close to the transition.

The period of the domain structure becomes exponentially large when the
modulus $\mu $ is anomalously small, $(D/\mu )^{1/2}\gg d_{at},$ and the
condition (\ref{eq26}) is violated. Then for very thin films $l\ll \left(
D/\mu \right) ^{1/2}$ one has to reconsider the calculations of the sum in
Eq.(\ref{eq20}). The exact result for $M=\mu$ $(\eta=1)$ is 
\begin{equation}
{\frac{\Delta F_{stray}}{{\cal A}}}={\frac{8\mu u_{0}^{2}a}{\pi ^{3}}}\left[ 
\frac{7}{8}\zeta (3)-Li_{3}(e^{-b})+\frac{1}{8}Li_{3}\left( e^{-2b}\right) %
\right] ,  \label{eq:Fstr0}
\end{equation}
where $b=2\pi l/a,$ $Li_{n}(z)\equiv \sum_{k=1}^{\infty }z^{k}/k^{n}$\cite
{Ryzhik}, cf. Refs. \cite{BLprl1,BLprl2}. The period of the domain
structure for a very thin film on a 
``soft substrate'' is exponentially large (cf. Refs. \cite{BLprl1,BLprl2}), 
\begin{equation}
a=\frac{\pi l}{e^{1/2}}\exp \left( \frac{\pi \gamma }{8\mu u_{0}^{2}l}%
\right) =1.9l\exp \left( \frac{\pi D^{1/2}}{3\mu ^{1/2}l}\right) .
\label{eq:alog}
\end{equation}

Next, we shall show that there is {\em no instability} with respect to the
structure with opposite domains of unequal widths, as suggested by Roytburd
\cite{Roytburd98}. To this end we shall estimate the change of the energy of
the system where the domains have different widths, $a(1+\delta )$ and $%
a(1-\delta )$. As we have discussed above, the homogeneous part of the
strain is zero. This is possible with domains of unequal width only if the
homogeneous strains in the domains, $u_{1}=u_{0}+\Delta u_{1}$ and $%
u_{2}=-u_{0}+\Delta u_{2},$ are different. The condition of zero homogeneous
strain reads 
\begin{equation}
u_{1}(1+\delta )+u_{2}(1-\delta )=0.  \label{eq:uconstr}
\end{equation}
We must also minimize the elastic energy of this structure. The energy
density of the homogeneous deformation can be found from the energy density
in the ferrophase, $2M(u_{xy}-u_{xy}^{0})^{2}+2\mu u_{xz}^{2}$ [cf. Eq.(\ref
{eq1})] which gives us the elastic energy of homogeneous stresses (cf. Ref.~
\cite{Roytburd98}) 
\begin{equation}
\frac{F_{h}}{{\cal A}}=\frac{1}{2}(1+\delta )2M\left( \Delta u_{1}\right)
^{2}+\frac{1}{2}(1-\delta )2M\left( \Delta u_{2}\right) ^{2}.  \label{eq:Fhd}
\end{equation}
Minimizing Eq.(\ref{eq:Fhd}) with respect to $\Delta u_{1}$ (which is
equivalent to the condition that the force acting on the domain wall is
zero), taking into account the constraint (\ref{eq:uconstr}), we obtain $%
\Delta u_{1}=\Delta u_{2}=-u_{0}\delta $ and 
\begin{equation}
F_{h}/{\cal A}=2Mu_{0}^{2}l\delta ^{2}.  \label{eq:Fhd2}
\end{equation}
One has to add the energy of inhomogeneous stresses, which depends on $%
\delta $%
\begin{equation}
\frac{\Delta F_{stray}\left( \delta \right) }{{\cal A}}=\frac{8\mu
u_{0}^{2}a\eta }{\pi ^{3}}\sum_{n=1}^{\infty }\frac{\left( -1\right)
^{n}(1-\cos \pi n\delta )}{n^{3}\left( \coth \frac{\pi n\eta l}{a}+\eta
\right) },  \label{eq:Fdel}
\end{equation}
where $\Delta F_{stray}\left( \delta \right) =F_{stray}\left( \delta \right)
-F_{stray}\left( 0\right) ,$ to obtain the total change in energy.

We shall now analyze two limiting cases.\newline
\noindent {\em Rigid substrate not very far from} $T_{c}$ ($\eta =\sqrt{%
M/\mu }\ll 1,$ $\pi \eta l/a\gg 1).-$ The expression in parentheses in the
denominator in Eq.(\ref{eq:Fdel}) can be replaced by unity, and we obtain 
\begin{eqnarray}
{\frac{\Delta F_{stray}}{{\cal A}}} &=&-{\frac{8\mu u_{0}^{2}a\eta }{\pi ^{3}%
}}\left[ \frac{3}{4}\zeta (3)+{\rm Re}Li_{3}(-e^{i\pi \delta })\right]  
\nonumber \\
&\approx &-{\frac{4\ln 2}{\pi }}\mu u_{0}^{2}a\eta \delta ^{2}.
\label{eq:Fsrigid}
\end{eqnarray}
In spite of the negative sign of this stray contribution, the total energy
increases with $\delta $, 
\begin{equation}
\frac{\Delta F}{{\cal A}}=2\eta \mu u_{0}^{2}a(\eta \frac{l}{a}-\frac{2\ln 2%
}{\pi })\delta ^{2}>0,  \label{eq:Fsrigd2}
\end{equation}
and, consequently, {\em no instability sets in}.

{\em Soft substrate} [$\left( D/\mu \right) ^{1/2}\gg d_{at},$ $\eta =1].-$
There the contribution (\ref{eq:Fdel}) can be found exactly with the
result \cite{Ryzhik}
\begin{eqnarray}
\frac{\Delta F_{stray}\left( \delta \right) }{{\cal A}} &=&-\frac{4}{\pi ^{3}%
}\mu u_{0}^{2}a\Bigl[\frac{3}{4}\zeta (3)+%
\mathop{\rm Re}%
Li_{3}\left( -e^{i\pi \delta }\right)   \nonumber \\
&&+Li_{3}\left( -e^{-b}\right) -%
\mathop{\rm Re}%
Li_{3}\left( -e^{i\pi \delta -b}\right) \Bigr].  \label{eq:Fssft}
\end{eqnarray}
In this case the domains may be wide, $l/a\ll 1,$ Eq.(\ref{eq:alog}), and we
obtain 
\begin{equation}
\frac{\Delta F}{{\cal A}}=\pi \mu u_{0}^{2}\frac{l^{2}}{a}\delta ^{2}.
\label{eq:Fssftd2}
\end{equation}
This energy increases with  $\delta$ and, therefore, the domain structure
{\em remains stable} on a ``soft'' substrate too.

Finally, we shall evaluate the {\em dielectric response} of the above domain
structure. We assume a linear coupling between $u_{xy}$ and $P_{z}$ given by
the thermodynamic potential

\begin{equation}
F=\int dV(2A_{1}u_{xy}^{2}+\frac{1}{2}A_{2}P^{2}+Bu_{xy}^{4}+gu_{xy}P)
\label{eq:FuP}
\end{equation}
with $A_{2}>0$, $P\equiv P_{z}$. Minimizing (\ref{eq:FuP}) with respect to $P
$ we obtain

\begin{equation}
A_{2}P+gu_{xy}=0,  \label{eq:Pu}
\end{equation}
hence the spontaneous strain $\pm u_{0}$ leads to the spontaneous
polarization in the domains $P_{0}=\mp gu_0/A_{2}.$ The previous free
energy functional (\ref{eq1}) can be obtained from (\ref{eq:FuP})\ with the
use of (\ref{eq:Pu}), and $A=A_{1}-g^{2}/4A_{2}.$

Let us show now that the net polarization {\em does not change} when the
domain walls have moved, so that now $\delta \neq 0.$ In a free standing
film the change would be $2P_{0}\delta$, but in a film on a substrate the
polarization itself changes since it follows the changes of the mean strains
in every domain. The homogeneous polarization $P_{h}$ is then, with the use
of Eq.~(\ref{eq:Pu}), 
\begin{eqnarray}
P_{h} &=&\frac{1}{2}\left( 1+\delta \right) P_{1}+\frac{1}{2}\left( 1-\delta
\right) P_{2}  \nonumber \\
&=&-\frac{g}{2A_{2}}\left[ \left( 1+\delta \right) u_{1}+\left( 1-\delta
\right) u_{2}\right] =0,  \label{eq:Ph}
\end{eqnarray}
because the average strain in the film (square brackets) is exactly zero,
Eq.~(\ref{eq:uconstr}). In other words in a thin film on a substrate the
shift of the domain walls does not lead to any change in the average
polarization. They appear to be {\em decoupled}, while they are strongly
coupled in a free standing film. Surprisingly, under an external field the
domain walls do not shift in our case and the field produces $P_{h}=E/A_{2}$%
: one has to put the field $E$ in Eq.$\left( \ref{eq:FuP}\right) $ and set $%
u_{xy}=0$ because the {\it homogeneous} deformation remains zero.

The result (\ref{eq:Ph}) emphasizes the qualitative difference in the
dielectric response of a ferroelectric-ferroelastic domain structure in a
free standing film and in an epitaxial film: (i) there is a finite stiffness
with respect to the domain wall shifts and (ii) these shifts change the
polarization within the domains for the film on a substrate. In our specific
case these changes completely compensate the polarization gain that would
have taken place in a free standing film. In general, the compensation is
not perfect but the effect should be taken into account. For example, these
two aspects have been overlooked by Erbil {\it et al}. \cite{Erbil} who have
only accounted for the energy change when the variation of the relative
domain fraction. This has been ascribed to a pinning potential, while the
other, much more important, terms were not considered. Their conclusion
about {\em enhancement} of the permittivity due to the domain structure is
incorrect. Our results, additionally, do not show any signs of instability
of the domain structure with the film thickness, as has been speculated
first by Roytburd\cite{Roytburd98} analytically and then by Pertsev {\em et
al.} \cite{Pertsev} numerically for perovskite ferroelectrics. Since their
case is somewhat different (the spontaneous strain $u_{xx}-u_{zz}$), it
might be interesting to apply their numerics to the present simpler case to
check the result.

In conclusion, we have demonstrated analytically how the domain structure
sets in at the phase transition and that it remains stable with respect to
spontaneous breaking of symmetry between the opposite domains irrespective
of whether the substrate is rigid or ``soft''. The dielectric response of
the epitaxial films is qualitatively different from that of free standing
films. In the present case the motion of the domain walls has {\em zero
effect }on dielectric response in ferroelastic-ferroelectric thin films.
Generally, the response of an epitaxial thin film is suppressed compared to
a free film.

\end{document}